\documentclass[fleqn,usenatbib]{mnras}

\usepackage{amsmath}
\usepackage{mathptmx}
\usepackage{txfonts}
\usepackage{graphicx}
\usepackage{adjustbox}
\usepackage{caption}
\usepackage{subcaption}
\usepackage[T1]{fontenc}

\DeclareRobustCommand{\VAN}[3]{#2}
\let\VANthebibliography\thebibliography
\def\thebibliography{\DeclareRobustCommand{\VAN}[3]{##3}\VANthebibliography}

\usepackage{graphicx}	
\usepackage{amsmath}	
\title[Probing the solar wind with Akatsuki]{On the estimation of solar wind velocity under varying solar activity conditions using Akatsuki measurements}

\author[Aggarwal et al.]{
Keshav Aggarwal$^{1}$\thanks{E-mail: keshavagg1098@gmail.com (KA)},
R. K. Choudhary$^{2}$,
Abhirup Datta$^{1}$,
T. Imamura$^{3}$
\\
$^{1}$Department of Astronomy, Astrophysics and Space Engineering (DAASE), Indian Institute of Technology Indore, Indore, Madhya Pradesh, 453552, India\\
$^{2}$Space Physics Laboratory (SPL), Indian Space Research Organization, Vikram Sarabhai Space Centre, Thiruvananthapuram, Kerala 695022, India\\
$^{3}$Graduate School of Frontier Sciences, The University of Tokyo, Kiban-tou 4H7, 5-1-5 Kashiwanoha, Kashiwa, Chiba 277-8561, Japan
}

\date{Accepted XXX. Received 2025 May 6; in original form ZZZ}

\pubyear{\the\year{}}

\begin{document}
\label{firstpage}
\pagerange{\pageref{firstpage}-\pageref{lastpage}}
\maketitle

\begin{abstract}
We present an analysis of solar wind dynamics based on Doppler spectral width measurements of X-band radio signals from the Japanese Akatsuki spacecraft. The dataset includes two solar conjunction occultation experiments conducted in 2016 and 2022, capturing the transition from the descending phase of Solar Cycle 24, a period of low solar activity, to the ascending phase of Solar Cycle 25, which exhibited moderate to intense activity. Our study demonstrates the utility of this technique for estimating both slow and fast solar wind velocities across different phases of solar activity. A key focus is the 2022 experiment, which probed the solar corona near coronal holes at heliocentric distances ranging from 1.4 to 10 $R_\odot$. We also investigate the impact of electron density estimates on the accuracy of solar wind speed determinations, underscoring the need for improved electron density modeling to enhance the robustness of such measurements.
\end{abstract}


\section{Introduction} 
\label{sec:intro}

The solar wind is driven by the extremely high temperatures of the coronal plasma and the influence of magnetic fields \citep{Parker1958, Cranmer2009}. As it propagates outward, the wind accelerates and ultimately reaches supersonic speeds \citep{West2023}. A crucial mechanism behind this acceleration is the conversion of magnetic energy into kinetic energy, facilitated by magnetohydrodynamic (MHD) waves and the reconfiguration of magnetic field lines into an "open" topology \citep{Cranmer2012}. These open field lines allow ionized particles to escape the Sun more freely, forming high-speed solar wind streams that can reach Earth in just three to four days \citep{Krieger1973, Nolte1976, Zirker1977}. When these fast streams interact with slower-moving solar wind, they generate Corotating Interaction Regions (CIRs). Although generally less intense than transient events like flares or Coronal Mass Ejections (CMEs), CIRs can still induce recurrent geomagnetic disturbances that persist across multiple Carrington rotations \citep{Bartels1934, Gosling1996, Balogh1999}.

Studies have shown that coronal heating is a major contributor to this in the near-Sun region of 2-10 $R_{\odot}$ \citep{West2023, Jain_2023, Jain_2024, Aggarwal2025}. While the high temperatures and magnetic dynamics in the solar corona are responsible for driving the solar wind, they also create distinct regions known as coronal holes where open magnetic field lines facilitate the rapid escape of plasma into space \citep{Cranmer2009}. Observations in extreme ultraviolet and soft X-ray wavelengths reveal these areas as dark patches, a signature of their lower plasma density and temperature relative to the adjacent closed-field regions \citep{Curdt2001, Aschwanden2013}. Their occurrence, evolution and latitudinal distribution are closely linked to the solar cycle; during solar minima, polar coronal holes are predominant, while low-latitude holes become more frequent near solar maximum \citep{Bravo1997}. These structures can persist for several solar rotations or evolve rapidly depending on underlying magnetic field dynamics.

Radio occultation (RO) has proven to be a powerful technique for probing the solar corona, especially in and around coronal holes. As radio signals traverse the turbulent coronal plasma, they experience scattering due to electron density fluctuations, which manifests as phase and intensity scintillations at Earth. Under weak-scattering conditions, these fluctuations carry signatures of the coronal plasma’s structure and dynamics \citep{Woo1976, Bird1982, Morabito2003, Chiba2022, Chiba2023}. By analyzing variations in the phase and amplitude of radio signals as they traverse the ionized coronal plasma, RO measurements provide detailed insights into the density structure and scale heights within these regions. This information is essential for understanding the mechanisms of solar wind acceleration and its broader interaction with the heliosphere. Over the decades, numerous missions have employed RO techniques using S- and X-band frequencies. Notable S-band studies include those conducted by Pioneer 6, Mariner 6/7, Pioneer 10/11, HELIOS 1 \& 2, Galileo, and the Mars Orbiter Mission \citep{Woo1976, Muhleman1977, Woo1978, Paetzold1987, Coles1991, Wohlmuth2001, Wexler2019, Jain2022, Jain_2024, Aggarwal2025}. Missions utilizing X-band signals include Cassini, Nozomi, MESSENGER, Akatsuki, and MAVEN \citep{Imamura2005, Morabito2007, Tokumaru2012, Miyamoto2014, Ando2015, Withers2018, Wexler2019, Wexler2020, Withers2020, Jain_2023, Jain_2024}. Dual-frequency RO experiments using both S- and X-band signals have been carried out by Viking, Ulysses, Mars Express, Rosetta, and Venus Express \citep{Tyler1977, Paetzold1996, Efimov2005}.

\begin{figure*}
\centering
    \includegraphics[width=\linewidth]{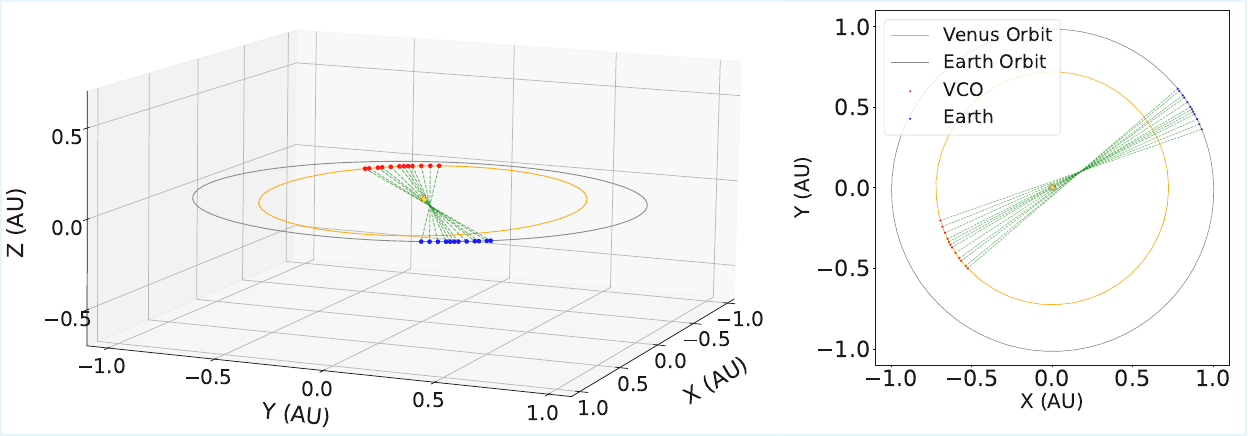}
    \caption{Schematic representation of Akatsuki-Venus-Sun-Earth geometry during radio occultation experiments. The blue points represent Earth's positions during 22-31 August 2022, while the red points represent positions of the Venus/ Venus Climate Orbiter (Akatsuki/VCO) with respect to the Sun and Earth during that period. The Left panel shows the geometry in the three-dimensional plane while the right panel shows its top-down view.}
    \label{fig:transit}

\end{figure*}

Using the X-band (8.41 GHz) signal from the Japanese Akatsuki spacecraft recorded during the 2016 and 2022 occultation events, we estimate solar wind velocities in the corona at varying heliocentric distances ranging from 1.4 to 10 $R_{\odot}$. We use the spectral characteristics of the radio signals for this purpose using a method described in \citet{Aggarwal2025}. The unique dataset\footnote{The Akatsuki team archives data from all radio science experiments, and this information is accessible to the public through their \href{https://darts.isas.jaxa.jp/missions/akatsuki/rs_en.html}{website}.} from Akatsuki, spanning a period of generally low solar activity, included instances of both slow and fast wind flows due to transient features such as equatorial coronal holes observed during the `Smiley-Sun' event which occurred during 22-31 October 2022. This coronal plasma region under investigation at heliocentric distances between 1.4 and 10 $R_\odot$ is crucial because it offers key observational data on the solar wind acceleration zone, the closest part of the Sun's atmosphere that has not been directly measured so far. The closest distance to Sun where the Parker Solar Probe could reach so far was $\sim$ 13.3 $R_\odot$ \citep{Badman2023}. In the lower corona, where magnetic fields dominate, soft X-ray and extreme ultraviolet instruments such as Hinode/XRT, PROBA 2/SWAP, and SDO/AIA have provided detailed studies of coronal loops, flares, and coronal holes, shedding light on plasma heating, magnetic field restructuring, and the sources of solar wind acceleration.

At heliocentric distances above the low corona ($\gtrsim$1.2-1.5 $R_{\odot}$), the solar wind acceleration region remains poorly understood due to limited in-situ coverage and the absence of a standardized remote sensing technique. This study aims to fill that gap by tracking radio signals from a satellite orbiting a planet, received on Earth while traversing regions of the solar corona that have remained unexplored due to technical limitations. The propagation of the spacecraft’s radio signal through this region of the corona enables remote sensing of local plasma conditions, such as electron density, turbulence, and solar wind velocity, by analyzing signal distortions introduced along the path. While a single LOS observation provides limited spatial coverage, the remote sensing technique demonstrated in this paper, when applied systematically across multiple occultation events, offers the potential to help fill the existing observational gap in the near-Sun solar wind, particularly in the 1.4-10 $R_\odot$ range, where in-situ data remain sparse. In addition to the quiet time conditions, this presents an opportunity to study even how coronal holes impact wind velocities in regions close to the Sun. These near-Sun measurements probing regions as close as 1.4 $R_\odot$ are invaluable for advancing our understanding of solar wind acceleration, its variability under different solar conditions, and for studying the extent to which coronal holes influence solar wind acceleration and plasma outflow near the Sun.  Figure \ref{fig:transit} shows a schematic of how the Akatsuki probe sent signals to Earth while in orbit around Venus, scanning through various regions of the solar corona during the October solar occultation event in 2022.

In this manuscript, section \ref{sec:maths} details the experimental setup and explains how spectral broadening in the received signal is calculated. Section \ref{sec:observations} demonstrates the scaling of the method developed in the previous study to make it applicable for X- band signals and derives solar wind speeds for both slow and fast wind regimes. These computed velocities are then compared with those from previous studies, and our conclusions are summarized in Section \ref{sec:ANSWER}.

\begin{table}
    \centering
    \caption{Akatsuki Mission Parameters.}
    \label{tab:mission_parameters}
    \begin{tabular}{lc}
        \hline
        Parameter & Value \\
        \hline
        \multicolumn{2}{c}{Mission Parameters} \\
        \hline
        Launch Date & 20 May 2010 \\
        Venusian Orbit Insertion & 7 December 2015 \\
        Planned Mission Duration & 2 years \\
        Duration of Operations & 13 years, 11 days \\
        Apoapsis & $\sim$ 360,000 km \\
        Periapsis & 400 km \\
        Orbital Period & $\sim$ 11 days \\
        Inclination & $\sim 3^{\circ}$ \\
        \hline
        \multicolumn{2}{c}{Antenna Parameters} \\
        \hline
        Diameter & 1.6m High Gain Antenna \\
        Operating Frequency & 8.41 GHz \\
        Power Requirement & 20 W DC Power \\
        Beamwidth & $\pm 2^{\circ}$ \\
        Peak Gain & 35 dB \\
        \hline
    \end{tabular}
\end{table}

\section{Methodology and Observations}
\label{sec:maths} 

Since its launch, \textit{Akatsuki} has been actively engaged in solar occultation experiments (see Table \ref{tab:mission_parameters} for mission details) \citep{Chiba2022, Chiba2023, Jain_2024}. In this study, we utilized the Venus solar conjunction events of 2016 (see Table \ref{tab:chiba2023}) and 2022 (see Table \ref{tab:combined}) to analyze the slow and fast solar wind conditions. The 2016 event corresponded to the descending phase of Solar Cycle 24, a period of low solar activity, while the 2022 event took place during the ascending phase of Solar Cycle 25, which exhibited moderate to intense activity.

Communications with Earth were conducted using the spacecraft's 1.6 m high-gain radial line slot antenna, equipped with oscillators exhibiting an Allan variance\footnote{For this experiment, the derived Allan deviations indicate frequency stability better than $10^{-12}$ over averaging times ranging from 1 to 1000 seconds.} on the order of $10^{-12}$. Telemetry, tracking, and command data were exchanged between the probe and the Usuda Deep Space Center, as detailed in \citet{Oshima2011, Imamura2017}. The transmitted X-band (8.41 GHz) signals were received by the 64-meter high-gain antenna at the Usuda Deep Space Center in Usuda, Nagano, Japan, using a radio science receiver \citep{Imamura2011}, and by the Indian Deep Space Network (IDSN) at Byalalu, India \citep{Jain_2024}.

The satellite radio signals were recorded using open-loop receiver systems and stored in the standard CCSDS-RDEF (Raw Data Exchange Format) in accordance with the Blue Book\footnote{\href{https://public.ccsds.org/Pubs/506x1b1.pdf}{https://public.ccsds.org/Pubs/506x1b1.pdf}}. The recorded signals, stored in the RDEF binary file, were processed using signal processing techniques involving signal extraction, Fourier transformation, and fitting of the Doppler power spectral density to estimate spectral broadening. These procedures are described in detail in \citet{Aggarwal2025}.

The unique geometry of the Akatsuki probe, specifically its highly elliptical orbit around Venus combined with favorable Earth-Sun-probe alignment during solar conjunctions, enabled us to investigate solar coronal holes at remarkably close distances, spanning approximately 1.4 to 10 $R_{\odot}$ (see Fig. \ref{fig:transit}). During conjunctions, the radio line-of-sight (LOS) from the probe to Earth pierced the corona at these offset distances, allowing us to remotely sense plasma properties along the ray path.

\begin{figure*}
\centering
 \includegraphics[width=\linewidth]{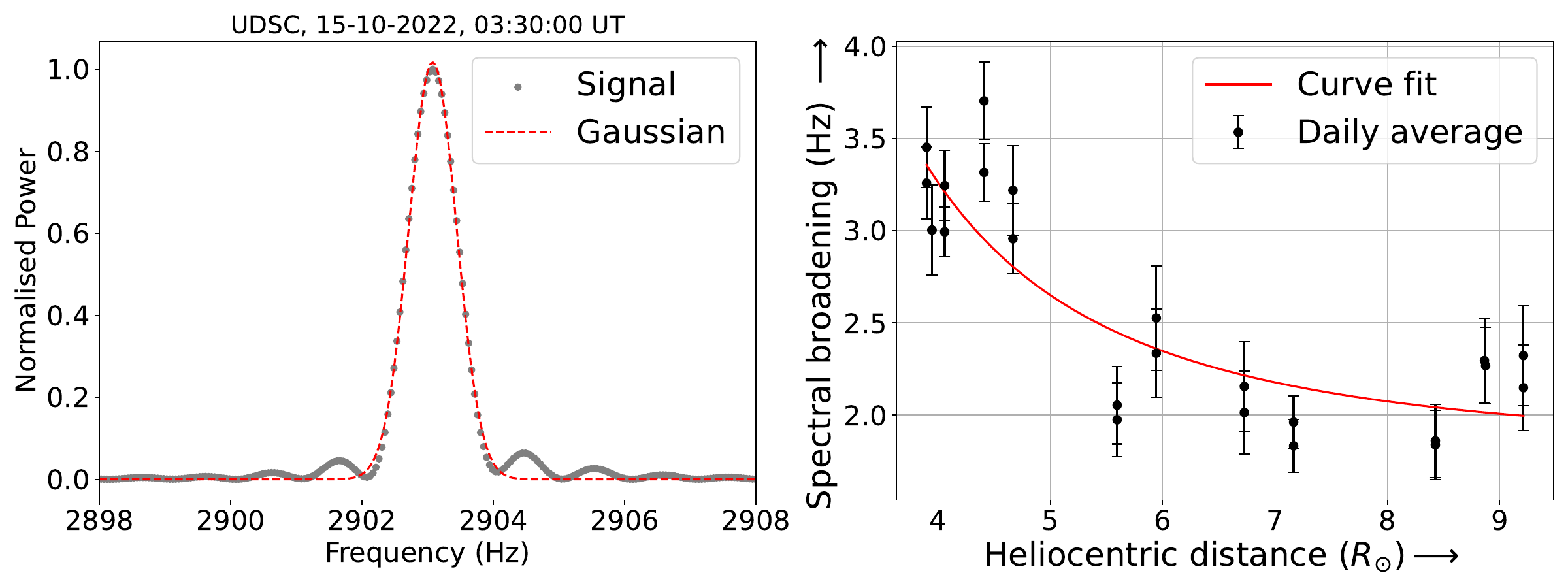}
 \caption{Left panel: Spectrogram for a 1 sec data sample with a Gaussian fit., Right panel: Daily average values of the spectral broadening as a function of increasing heliocentric distance for the observations in 2022.}
\label{fig:spectrum}
\end{figure*}

\subsection{Data processing}
\label{sec:dataprocessing}
Fig. \ref{fig:spectrum} (left panel) shows a Gaussian fit to a sample spectrogram of the received signal for a 1-s data frame at 03:30:00 UT. The gray points represent the spectral power density, while the red line is the Gaussian model fit. Following the method described in \citet{Aggarwal2025}, we assume that the power spectrum of the signal is approximately Gaussian in shape, allowing us to estimate key physical parameters from the zeroth, first, and second moments of the distribution. The total power ($P$), Doppler shift ($\Omega$), and spectral width ($B_S$) are extracted using statistical estimators. The spectral width parameter provides a measure of Doppler broadening induced by fluctuations in the coronal plasma arising from both electron density turbulence and velocity dispersion in the solar wind. This broadening serves as the basis for estimating the bulk plasma velocity component perpendicular to the LOS. This approach assumes that the coronal plasma acts as a distributed weak-scattering medium and that the spectral shape of the received signal is not significantly distorted by instrumental effects.

To reduce fluctuations in the data, we applied a one-minute moving average, as the solar offset distances remain relatively constant over this interval \citep{Woo1979, Coles1991}. Each occultation observation typically spans approximately 6.5 hours, during which the solar offset of the radio LOS changes by only about $\sim$0.1$R_{\odot}$. Given this limited variation, we compute a daily average of the spectral broadening values for each observation. Figure \ref{fig:spectrum} (right panel) presents these daily mean broadening values as a function of the corresponding average heliocentric distance.

Considering the oscillator's Allan variance of $10^{-12}$, the associated error due to signal instability is calculated as $10^{-12} \times f = 0.00841$ Hz \citep{Tripathi2022b}. In comparison, the standard deviation of the spectral broadening measurements typically ranges from 15-30\% of the absolute value. However, this impact is mitigated when using $B_X^{1/6}$ for velocity estimation. After propagation through the empirical formulation, the resulting uncertainty in $B_X^{1/6}$ averages to approximately 0.035 Hz--about four times larger than the contribution from oscillator instability. This indicates that instrumental instability is negligible relative to the statistical variability in the measurements.

The total uncertainty in spectral broadening is computed by combining the oscillator instability error and the statistical standard deviation of the daily measurements in quadrature. This approach yields a more realistic estimate, assuming that both contributions are independent and Gaussian-distributed.

\subsection{Spectral width and solar wind velocity}
\label{sec:methods}

An empirical relationship was derived in \citet{Aggarwal2025} to relate Doppler spectral width to the bulk solar wind velocity component perpendicular to the LOS, based on radio occultation observations and validated for S-band signals. A simplified formulation was developed to directly relate Doppler spectral width to the solar wind velocity, as shown in Eq. (\ref{eq:method}). The derived solar wind speeds ranged from $100$-$150\,\mathrm{km\,s^{-1}}$. The estimated electron densities ($\sim10^{10}\,\mathrm{m^{-3}}$) were consistent with previous models but were on the lower end, likely due to the low solar activity during the observation period \citep{Edenhofer1977, Esposito1980, Muhleman1981, Strachan1993, Guhathakurta1994, Cranmer1999, Wexler2019}.

As established in that study, the solar wind bulk velocity component perpendicular to the LOS can be estimated from the observed spectral broadening of the received signal using the following empirical relation:
\begin{equation}
\label{eq:method}
v_{\perp} = k_0 \left[ \frac{r \times R_{\mathrm{EP}} \times (1+R_{\mathrm{SP}})^2}{R_{\mathrm{SP}}} \right] B_S^{\frac{1}{6}}
\end{equation}
where \(v_{\perp}\) is the solar wind speed component perpendicular to the LOS, \(B_S\) is the spectral broadening (in Hz), and \(r\) is the solar offset distance—defined as the distance of closest approach to the Sun along the LOS. The constant \(k_0 = 1.687\); \(R_{\mathrm{EP}}\) is the distance from the probe to Earth (with 1 AU equal to \(1.496 \times 10^{11}\) meters); and \(R_{\mathrm{SP}}\) is the distance between the Sun and the probe, expressed in AU.

This formulation assumes a Kolmogorov-type turbulence spectrum in the coronal plasma and a steady-state solar wind outflow. At the heliocentric distances probed in this study (1.4–10 $R_{\odot}$), it is further assumed that the solar wind flows nearly radially outward and that the coronal magnetic field is predominantly radial \citep{Woo1977, Waldmeier1977}. Under these conditions, the perpendicular velocity component \(v_{\perp}\) provides a good approximation of the total solar wind bulk flow speed.

The advantage of this approach lies in its simplicity and ease of use with Doppler spectrograms obtained during radio occultation. It enables quick and reliable extraction of solar wind speeds from observational data because it depends on directly observable quantities, such as spectral broadening, has an analytical form with few free parameters, and has shown consistency with independent velocity estimates derived from phase scintillation and frequency residual techniques  \citep{Jain_2024, Aggarwal2025}.

Although the initial method was designed for S-band radio signals, adjustments are needed for X-band frequencies (8.4 GHz) to account for lower scattering cross-section and narrower Doppler widths at higher frequencies. The frequency-dependent scattering differences require modifying the empirical constants when transitioning from S-band to X-band. These changes and their validation using Akatsuki are detailed in subsequent sections. Additional theoretical derivation and validation are provided in  \citep{Aggarwal2025}, which offers a comprehensive analysis of Doppler spectral techniques, turbulence modeling, and solar wind parameter estimation from radio occultation data. The methodology outlined in  \citep{Aggarwal2025} is adopted here for X-band analysis which is as follows :

\begin{enumerate}
    \item Calculate the spectral broadening in the received signal and the point of closest approach to the Sun.
    \item Assume the density fluctuation spectrum follows a single power law with spectral index $p = 11/3$ (Kolmogorov-type turbulence).
    \item Estimate electron density $N_e$ using the empirical relation calibrated for near-Sun RO observations.
    \item Derive solar wind velocity $v_\perp$ perpendicular to the radio LOS, using the measured spectral bandwidth and estimated $N_e$.
\end{enumerate}

\begin{table}
\centering
\caption{Parameters and constants used in the empirical equations, along with their units and values where applicable.}
\label{tab:parameters}
\setlength{\tabcolsep}{4pt} 
\begin{tabular}{lccc}
\hline
Symbol & Meaning & Value & Unit \\
\hline
$f$ & Signal frequency (X-band) & 8.41 & GHz \\
$\lambda$ & Signal wavelength & 3.57 & cm \\
$B_X$ & Spectral broadening & Measured & Hz \\
$N_e$ & Electron density & Derived & m$^{-3}$ \\
$v_\perp$ & Solar wind speed & Derived & km s$^{-1}$ \\
&(perpendicular to LOS)&&\\
$R_{EP}$ & Earth-Probe distance & Variable & AU \\
$R_{SP}$ & Sun-Probe distance & Variable & AU \\
ESP & Earth-Sun-Probe angle & Variable & radians \\
$r_e$ & Classical electron radius & $2.8179 \times 10^{-15}$ & m \\
$c_0$ & Empirical constant & $1.14 \times 10^{-24}$ & -\\
$k_0$ & Scaling constant (S-band) & 1.687 & - \\
AU & Astronomical unit & $1.496 \times 10^{11}$ & m \\
Allan var. & Oscillator stability (1-1000 s) & $\sim10^{-12}$ & -\\
\hline
\end{tabular}
\end{table}

Table \ref{tab:parameters} summarizes all the key parameters and constants used throughout the empirical formulations, including their units and, where applicable, numerical values.

The dataset analyzed in this study comprises a total of 19 observation days: 8 days from the 2016 solar conjunction experiment and 11 days from the 2022 event. Each day corresponds to a distinct radio LOS path through the solar corona, resulting in 19 independent LOS geometries spanning heliocentric distances from 1.4 to 10 $R_{\odot}$. These individual observations form the basis for the derived solar wind velocity and density estimates reported in the following sections.

\section{Results and Discussion}
\label{sec:observations}
\subsection{Slow solar wind measurements}
The observations conducted with Akatsuki between 30 May and 15 June 2016 coincided with the spacecraft’s solar conjunction, as viewed from the ground-based receiver on Earth. This period provided a valuable opportunity to probe the near-Sun slow solar wind using the RO technique, enabling the retrieval of slow solar wind velocities with our method. Using the same dataset, \citet{Chiba2022}, through interplanetary scintillation (IPS) maps, heliospheric magnetic field extrapolations, and coronal imaging, confirmed that, with the exception of 30 May, 1 June, and 3 June, the observed regions were disconnected from coronal holes and did not exhibit high-speed solar wind flows. The remaining days (particularly 4, 5, and 8–15 June) were therefore identified as being dominated by slow wind conditions, primarily at heliographic latitudes between $-2.6^\circ$ and $+0.74^\circ$. Supporting evidence from white light and EUV images obtained by Hinode/XRT, along with magnetic potential field maps (e.g., Fig. 3 in \citealt{Chiba2022}), further confirmed the absence of open-field coronal hole structures along the line of sight on these dates. Additionally, WSA-Enlil model\footnote {\href{https://www.swpc.noaa.gov/products/wsa-enlil-solar-wind-prediction}{https://www.swpc.noaa.gov/products/wsa-enlil-solar-wind-prediction}} predictions for the same period  also did not indicate any high-speed streams at the sampled latitudes. The heliocentric distances probed ranged from 1.36 to 9.00 $R_\odot$ and, except for 9 June, were not influenced by transient events such as coronal mass ejections (CMEs).

To adapt our spectral width analysis for X-band signals, we applied a scaling correction based on the theoretical relationship between bandwidth broadening ($B$), wavelength ($\lambda$), and the turbulence spectral index. For a Kolmogorov spectrum with $p = 11/3$, the spectral width ratio between S- and X-band signals scales as:

\begin{equation}
\label{eq:freqbroad}
    \frac{B_X}{B_S} = \left( \frac{\lambda_S}{\lambda_X} \right)^{\frac{2}{p - 2}} = \left( \frac{\lambda_S}{\lambda_X} \right)^{\frac{6}{5}} \approx 0.2,
\end{equation}

This empirical relation assumes that the power spectral density of coronal turbulence follows a Kolmogorov spectrum.  This is a standard assumption for weak-scattering radio occultation studies in the inner corona and has been widely adopted in previous analyses \citep{Woo1976, Bird1982, Morabito2003, Aggarwal2025}. The Kolmogorov spectrum represents fully developed inertial range turbulence and provides a good fit to observed radio signal fluctuations at heliocentric distances of 1.4-10 \(R_{\odot}\). This implies that a S-band signal will show five times more broadening than an X-band signal in the same conditions \citep{Morabito2003, Morabito2009}. Hence, to apply the methodology described in Section \ref{sec:methods} to X-band data, the observed spectral broadening must be scaled by a factor of 5, making the final equation for slow solar wind speeds:

\begin{equation*}
    v_{\perp} =k_0 \left[ \frac{r\times  R_{EP} \times (1+R_{SP})^2}{R_{SP}} \right] (5B_X)^{\frac{1}{6}} 
\end{equation*}
\begin{equation}
\label{eq:slowwind}
\phantom{v_{\perp} }= 1.3 \times k_0 \left[ \frac{r\times  R_{EP} \times (1+R_{SP})^2}{R_{SP}} \right] B_X^{\frac{1}{6}}
\end{equation}

As previously mentioned, this event was also analyzed by \citet{Chiba2022} using X‑band downlink signals from \textit{Akatsuki}. They employed intensity scintillations in the radio signals to probe coronal plasma fluctuations through weak‑scattering theory. As the spacecraft’s radio signal traverses the turbulent coronal plasma, it undergoes scattering, producing a moving diffraction pattern at Earth due to solar wind advection. By fitting theoretical power spectra to the observed intensity fluctuations, they retrieved solar wind velocities and other plasma parameters.

The theoretical model adopted by \citet{Chiba2022} incorporated key turbulence characteristics, including spectral slope, axial ratio, and inner dissipation scale, treating the spectral index as a free parameter in their fits to the scintillation spectra. Their analysis also accounted for Fresnel filtering and turbulence anisotropy, and assumed a thin scattering screen located at the point of closest solar approach. However, the method is valid only under weak‑scattering conditions and loses accuracy within approximately \(3\,R_{\odot}\) due to the breakdown of its underlying assumptions. Additionally, the LOS integrated nature of scintillation measurements limits spatial resolution along the radial direction. Although the model assumes a localized thin screen, in practice the coronal plasma perturbs the signal over a finite path length along the LOS. The resulting power spectrum therefore represents an integrated response to density fluctuations across this distributed segment, smoothing out spatial variations. This integration makes it challenging to isolate plasma properties, such as \(v\), \(N_e\), or turbulence parameters, at a specific heliocentric distance, thereby reducing the diagnostic precision.

\begin{figure}
\centering
\includegraphics[width=\linewidth]{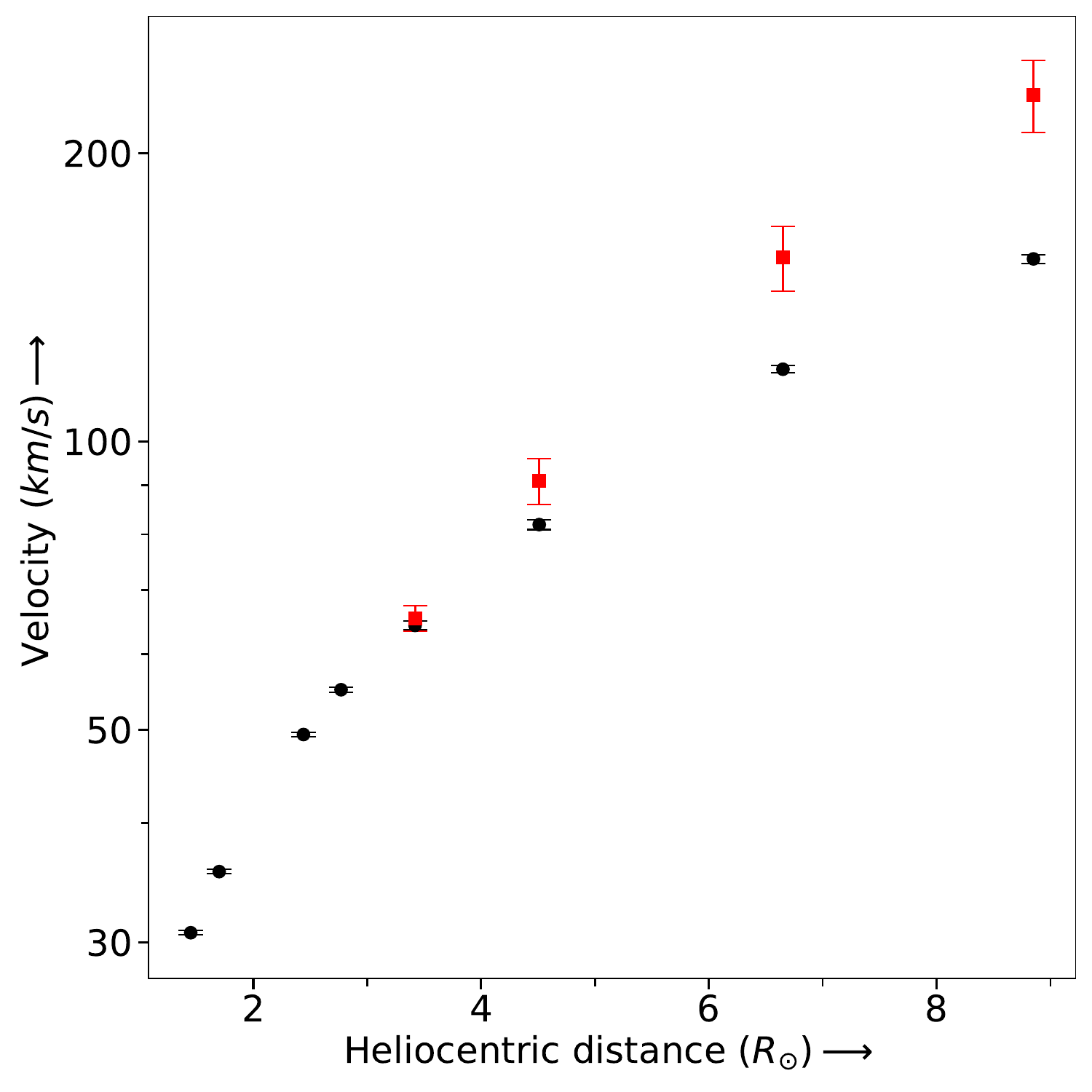}
\caption{Our spectral width method applied to the Akatsuki X-band observations in 2016. The black points represent our derived slow solar wind velocities with associated error bars. The red points are velocity estimates from \citet{Chiba2022}, obtained using intensity scintillation analysis. Both sets of results are based on the same observation period during the 2016 solar conjunction, but derived using distinct methodologies: spectral broadening versus weak-scattering scintillation modeling.}
\label{fig:chiba}
\end{figure}

We employed the spectral broadening approach for the detailed mapping of slow and fast solar wind streams in the $1.4$-$10R_{\odot}$ range, which revealed distinct turbulence properties between the two regimes. As shown in Table \ref{tab:chiba2023} and Fig. \ref{fig:chiba}, the derived solar wind velocities obtained using our scaled spectral broadening method exhibit good agreement with independent estimates from the same Akatsuki dataset. Although not all values strictly overlap within 1-$\sigma$ uncertainties, the relative agreement between our velocity estimates and those from \citet{Chiba2022}, combined with the substantially smaller and consistently low uncertainties from our method, supports the robustness of the spectral broadening technique at X-band frequencies. The error magnitudes remain of the same order across multiple independent observations, reinforcing their reliability for solar wind diagnostics.

\begin{table}
    \centering
    \caption{Slow solar wind velocities derived from the Akatsuki solar occultation experiment conducted in 2016 using our spectral broadening method. For comparison, values from \citet{Chiba2022} obtained using intensity scintillation analysis of the same dataset are also listed. The final column shows the percent difference between the two methods wherever comparison is possible.}
    \setlength{\tabcolsep}{2.5pt} 
    \label{tab:chiba2023}
    \begin{tabular}{lcccccc}
        \hline
        & Heliocentric  & Chiba et al. & Our  $V$ & $V_{\text{err}}$ & $V_{\text{err}}$ & $\Delta V$ \\
        Date & distance ($R_{\odot}$) & (km/s) &(km/s) & (km/s) & (\%) & (\% diff) \\
        \hline
        04-Jun-2016 & 2.77 & -     & 55.08  & $\pm$0.34 & $\pm$0.61 & - \\
        05-Jun-2016 & 1.70 & -     & 35.58  & $\pm$0.18 & $\pm$0.50 & - \\
        08-Jun-2016 & 1.45 & -     & 30.72  & $\pm$0.15 & $\pm$0.50 & - \\
        09-Jun-2016 & 2.44 & -     & 49.46  & $\pm$0.27 & $\pm$0.54 & - \\
        10-Jun-2016 & 3.42 & 65.35  & 64.28  & $\pm$0.66 & $\pm$1.03 & 1.63 \\
        11-Jun-2016 & 4.51 & 91.00  & 81.90  & $\pm$0.98 & $\pm$1.19 & 10.00 \\
        13-Jun-2016 & 6.65 & 155.55 & 118.96 & $\pm$1.09 & $\pm$0.92 & 23.53 \\
        15-Jun-2016 & 8.85 & 230.00 & 155.07 & $\pm$1.61 & $\pm$1.04 & 32.59\\
        \hline
    \end{tabular}
\end{table}

\begin{figure*}
\centering
\includegraphics[scale=0.4]{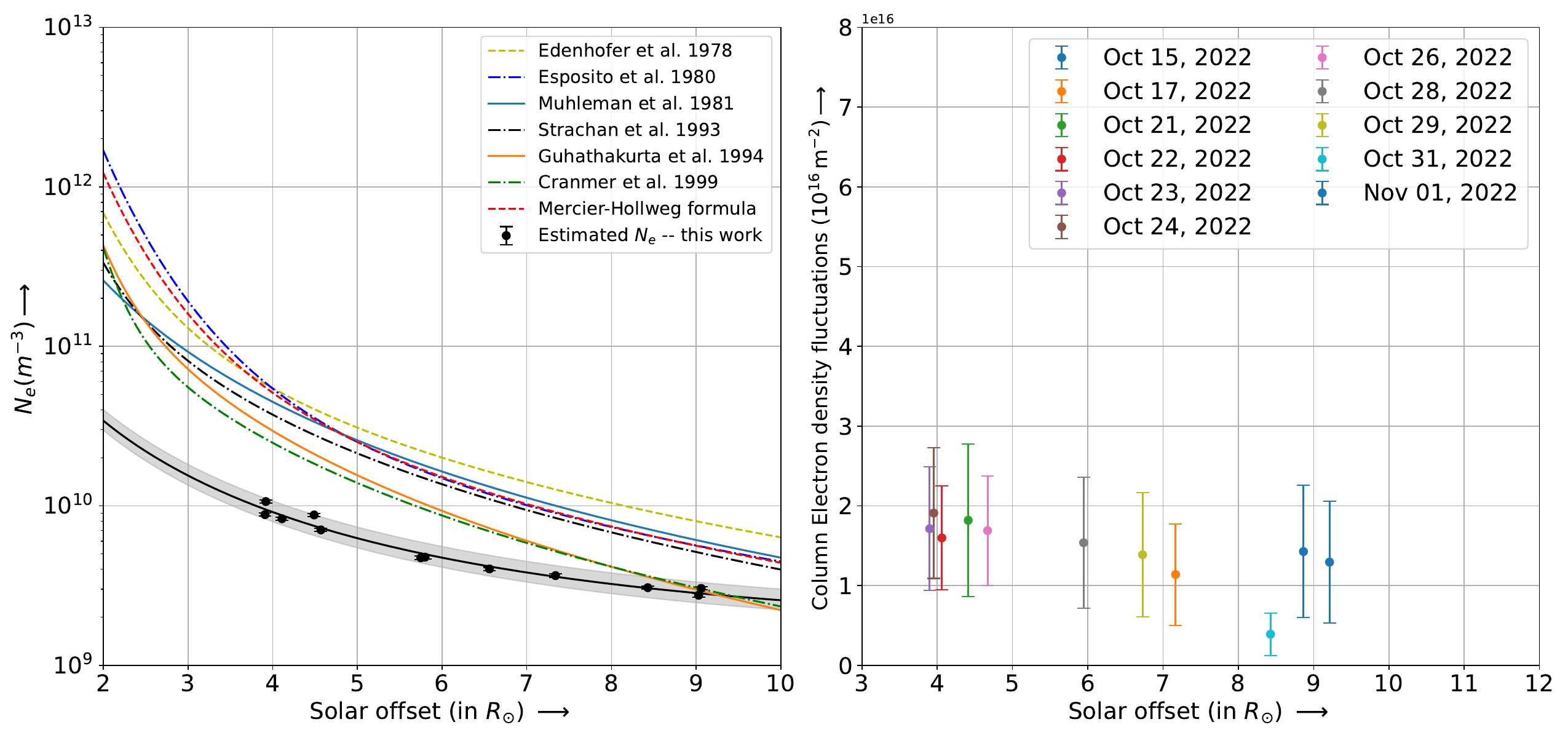}
\caption{Left panel: Electron density estimates for the 2022 solar occultation experiment compared against other models in the literature. Right panel: Fluctuations in the column density during October 2022.}
\label{fig:densities_2}
\end{figure*}

\subsection{Solar wind speeds near coronal holes}
Among the various contributors to spectral broadening such as turbulence, thermal motion, LOS Doppler drift, and instrumental effects; fluctuations in electron density due to coronal turbulence are dominant in our observations \citep{Bird1982, Morabito2003}. For heliocentric distances of 1.4-10 $R_{\odot}$ at X-band frequencies, turbulence induced broadening typically ranges from 0.1-1.0 Hz, while thermal broadening contributes $\lesssim$0.001 Hz (assuming $T_e \sim 1$ MK), and oscillator instability adds $\sim$0.008 Hz. The electron densities near the Sun using X-band signals can be roughly estimated using the following empirical relation, assuming that the density spectrum has a Kolmogorov spectral index of $ p=11/3 $, a spherically symmetric coronal density, and a steady-state outflow. Using Eq. (\ref{eq:freqbroad}), and \citet{Woo1977, Woo1978, Bird1990, Ho2002, Morabito2003, Yunqiu2015, Aggarwal2025}, the electron densities at the point of closest approach can be roughly estimated using the broadening in the X-band signal as:
\begin{equation}
    N_e = \frac{ f }{r\times \text{[ESP]} } \times \left( \frac{5B_X}{c_0 } \right)^{ \left( \frac{5}{6} \right)}
    \label{eq:N_e}
\end{equation}
Here, \( f \) is the signal frequency in GHz, \([\text{ESP}]\) is the Earth-Sun-Probe angle in radians, and \( c_0 = 1.14 \times 10^{-24} \), providing density estimates in m\(^{-3}\). This empirical relationship, derived using X-band radio signals from spacecraft in near-ecliptic orbits, yields reliable electron density estimates when applied to observations within or near the ecliptic plane—regions typically associated with streamer belts and the slow solar wind \citep{Aggarwal2025}.

However, earlier studies have shown that electron densities in coronal hole regions are lower by a factor of approximately 2-2.7 compared to those in the ecliptic plane \citep{Doschek1997, Gallagher1999}. By incorporating this reduction factor, we can extend the empirical relation to estimate electron densities in coronal hole regions, thereby providing approximate values for plasma densities close to the Sun.

In the left panel of Fig. \ref{fig:densities_2}, we present our electron density estimates for the 2022 Akatsuki experiment. The shaded envelope in grey represents the range of density reduction factors reported in earlier studies \citep{Doschek1997, Gallagher1999}. As evident from the figure, our density estimates are significantly lower than those predicted by other empirical models in the literature \citep{Edenhofer1977, Esposito1980, Muhleman1981, Strachan1993, Guhathakurta1994, Cranmer1999, Wexler2019}.

To represent the difference in densities between the ecliptic and coronal hole regions, we adopt an average reduction factor of \( \sim 2.35 \), based on the range reported in the aforementioned studies. From \citet{Woo1977, Woo1978, Waldmeier1977, Aggarwal2025}, we obtain \( v_\perp \), the solar wind bulk velocity component perpendicular to the LOS, as:
\begin{equation*}
    v_{\perp} = \frac{1}{N_e} \times \frac{11.54 \times B \times R_{EP} (1+ R_{SP})}{k \times R_{SP}^2 \times r_e \times \lambda^2}
\end{equation*}
or more simply as 
\begin{equation}
    v_{\perp}  = \frac{1}{N_e} \times [\text{Other parameters}]
\end{equation}

Here, the "other parameters" refer to geometric and instrumental constants which are independent of the solar wind regime (e.g., slow or fast wind). Which then gives us 

\begin{equation}
    \frac{( v_{\perp})_{\text{ecliptic}}}{( v_{\perp})_{\text{coronal holes}}} = \frac{( \frac{1}{N_e})_{\text{ecliptic}}}{( \frac{1}{N_e})_{\text{coronal holes}}} = \frac{(N_e)_{\text{coronal holes}}}{(N_e)_{\text{ecliptic}}} \sim 2.35
\end{equation}
Here, the values of \(N_e\) used in the ratio are not directly measured from our data but are based on typical empirical reduction factors reported in earlier studies comparing electron densities in coronal holes and equatorial streamer regions \citep{Doschek1997, Gallagher1999}. We adopt a representative factor of 2.35 to quantify this density contrast. Accordingly, the velocities obtained from Eq. (\ref{eq:slowwind}) require a correction factor of approximately 2.35 when applied to X-band data probing coronal holes, where electron densities are substantially lower than in the ambient equatorial solar wind. This scaling reflects an empirical relationship between density and inferred wind speed, rather than a strict conversion between different solar wind regimes.

While this density-based correction plausibly accounts for the lower velocities observed in coronal holes, other factors, particularly the turbulence properties, may also influence the results. The inferred velocities are sensitive to the assumed turbulence spectrum. Eq. (\ref{eq:slowwind}) is derived under the assumption of a Kolmogorov spectral index (\(p = 11/3\)), but several studies have reported flatter spectra (e.g., \(p \sim 3\)) in the inner corona, especially between 6-10 \(R_{\odot}\) and within coronal hole regions \citep{Woo1979, Kolosov1982, Morabito2003, Jain_2024}. Such deviations could alter the relationship between spectral broadening and solar wind velocity, introducing additional uncertainty. Future work will examine the sensitivity of our velocity estimates to variations in \(p\), which may provide further insight into the turbulence regimes of different coronal environments.

We computed density fluctuations at a cadence of 1 Hz and report average values over the full observation period. The right panel of Fig. \ref{fig:densities_2} presents the daily mean fluctuations as a function of heliocentric distance. Most days exhibit a gradual trend consistent with the distance-dependent evolution of turbulence. However, the measurement on October 31 appears anomalously low. To investigate this, we examined the Akatsuki - Earth LOS geometry and consulted heliospheric models. EUV imaging and a prior study by \citet{Jain_2024} indicate that the LOS continued to remain embedded in the Solar open field lines throughout the period of observation.  Nevertheless, no clear instrumental or geometric anomaly was identified that could explain the abrupt drop in column density fluctuations on that day. Notably, \citet{Jain_2024} also excluded October 31 from their published velocity time series, possibly reflecting similar uncertainty. We therefore refrain from drawing definitive conclusions for this date and encourage future studies involving coronal magnetic field extrapolations or PFSS modeling to explore the anomaly further.

\citet{Jain_2024} also studied the ``Smiley-Sun'' event, employing the coronal radio-sounding technique using one-way X-band downlink signals from the Akatsuki spacecraft. Their method, however, relied on the premise that radio signals traversing the turbulent inner corona undergo Doppler frequency fluctuations due to variations in electron density. By subtracting the modeled geometric Doppler shift-calculated from spacecraft ephemerides-from the observed signal, they obtained the residual frequency fluctuations. Under the assumption that these residuals encoded the time-dependent power spectrum of the coronal plasma, they applied Fourier transforms to the residuals to determine the spectral slope, which was later used, in conjunction with Taylor's frozen-in turbulence hypothesis, to estimate the bulk solar wind velocity along the radio ray path.

While effective for probing near-Sun regions (3-9 $R_{\odot}$), the method relied on several assumptions: isotropic and quasi-stationary turbulence, a direct link between temporal and spatial scales, and empirical background inputs such as electron density models and turbulence scaling laws. Furthermore, because the technique integrated plasma effects along the entire LOS, it had limited spatial resolution. Despite these limitations, it remains a valuable tool for remotely diagnosing inner coronal dynamics--regions largely inaccessible to \textit{in-situ} measurements.

\begin{table*}
    \centering
    \caption{Derived slow and fast solar wind velocities and associated uncertainties from the Akatsuki probe for the October 2022 solar occultation event (in km/s), using our spectral broadening method \citep{Aggarwal2025}. For comparison, estimates from \citep{Jain_2024}, which are based on Doppler frequency fluctuation analysis, are also included.}
    \label{tab:combined}
    \setlength{\tabcolsep}{4pt}
    \begin{tabular}{lcccccccc}
        \hline
         & Heliocentric &  & $V_{\text{slow}}$ &  & $V_{\text{fast}}$ & \cite{Jain_2024} & \cite{Jain_2024} & Difference \\ 
        Date & Distance ($R_{\odot}$) & $V_{\text{slow}}$ (km/s) & Error (km/s) & $V_{\text{fast}}$ (km/s) & Error (km/s) & (km/s) & Error  (km/s) & (\%) \\ \hline
        15-Oct-2022 & 9.03 & 149.94 & $\pm$0.84 & 352.36 & $\pm$1.98 & 381.74 & 26.97 & 7.70 \\ 
        17-Oct-2022 & 7.34 & 122.31 & $\pm$1.18 & 287.43 & $\pm$2.77 & 320.97 & 17.48 & 10.45 \\ 
        19-Oct-2022 & 5.75 & 96.24 & $\pm$0.98 & 226.17 & $\pm$2.31 & 256.55 & 14.48 & 11.84 \\ 
        21-Oct-2022 & 4.49 & 80.04 & $\pm$1.24 & 188.09 & $\pm$2.90 & 208.6 & 12.6 & 9.83 \\ 
        22-Oct-2022 & 4.11 & 71.89 & $\pm$1.12 & 168.93 & $\pm$2.63 & 190.74 & 14.65 & 11.43 \\ 
        23-Oct-2022 & 3.91 & 68.83 & $\pm$1.03 & 161.75 & $\pm$2.42 & 182.63 & 9.69 & 11.43 \\ 
        24-Oct-2022 & 3.92 & 72.09 & $\pm$1.26 & 169.42 & $\pm$2.96 & 183.39 & 12.6 & 7.62 \\ 
        26-Oct-2022 & 4.57 & 82.15 & $\pm$1.31 & 193.05 & $\pm$3.08 & 209.96 & 15.22 & 8.05 \\
        28-Oct-2022 & 5.8 & 102.2 & $\pm$1.31 & 240.16 & $\pm$3.08 & 258.67 & 21.26 & 7.16 \\ 
        29-Oct-2022 & 6.56 & 114.79 & $\pm$1.06 & 269.75 & $\pm$2.49 & 287.89 & 20.39 & 6.30 \\ 
        31-Oct-2022 & 8.43 & 143.01 & $\pm$1.27 & 336.08 & $\pm$2.97 & - & - & - \\ 
        01-Nov-2022 & 9.06 & 157.61 & $\pm$1.50 & 370.38 & $\pm$3.53 & 383.29 & 24.03 & 3.37 \\ 

        \hline
    \end{tabular}
\end{table*}

\begin{figure}
\centering
\includegraphics[width=\linewidth]{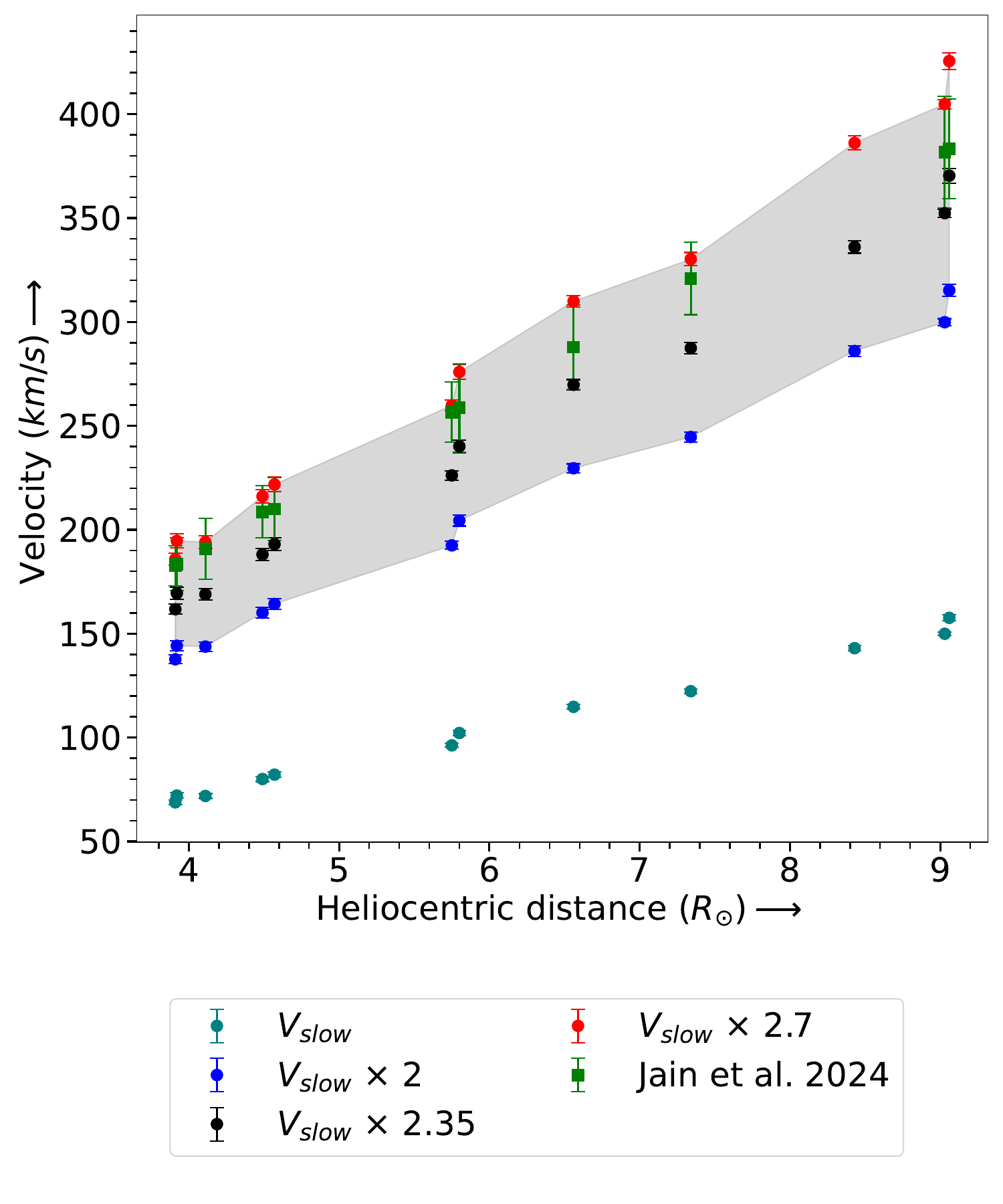}
\caption{Our method applied to the Akatsuki X-band observations in 2022. $V_{\text{slow}}$ denotes the wind velocities when the scaling for coronal hole densities is not done. $V_{\text{slow}} \times 2$ incorporates the density scaling by \citet{Doschek1997}, $V_{\text{slow}} \times 2.7$ shows the scaling by \citet{Gallagher1999}, while $V_{\text{slow}} \times 2.35$ considers the average of the two. The Grey shaded region is an envelope which shows the variation in velocities for scaling of $2-2.7$.}
\label{fig:history}
\end{figure}

In Fig. \ref{fig:history} we present the derived solar wind velocities obtained from the spectral broadening of the received X-band signals, as computed using our empirical formulation (Eq. \ref{eq:slowwind}). These results are compared with independent estimates reported in the literature, particularly those by \citet{Jain_2024}. Table \ref{tab:combined} lists the corresponding bulk velocity component perpendicular to the LOS estimates derived from the radio occultation (RO) observations conducted in October 2022 using the \textit{Akatsuki} spacecraft. These observations targeted regions near the equatorial coronal holes, offering a valuable opportunity to test the method under high-speed solar wind conditions. Each LOS during the October 15-November 1 occultation event intersected or passed close to the equatorial coronal holes identified during the well-known “Smiley-Sun” configuration, as described in \citet{Jain_2024}. Although we do not employ WSA-Enlil or PFSS modeling in this study, qualitative assessment from EUV imagery suggests that all LOS paths sampled coronal hole-associated fast wind regions rather than ambient equatorial streamers. Future work may incorporate full coronal magnetic mapping to further refine LOS-region classifications.

The comparison (Fig. \ref{fig:history}) reveals that our derived solar wind velocities during the 2022 solar conjunction closely track the estimates by \citet{Jain_2024}, who employed phase scintillation analysis of Akatsuki X-band signals to characterize the high-speed wind streams originating from equatorial coronal holes. The agreement between the two independent analyses, despite differing methodologies, reinforces the robustness of radio occultation as a remote sensing tool for coronal plasma dynamics. These results also show the applicability of our empirical formulation (Eq. \ref{eq:slowwind}) in capturing solar wind acceleration signatures across varying heliocentric distances and coronal conditions, including high-speed flows associated with equatorial hole regions.

While the agreement with independent methods and the consistency across multiple days reinforce the validity of our approach, we emphasize a few key limitations. The method employed in \cite{Jain_2024} relies on an empirical formulation that assumes Kolmogorov-type turbulence and uses electron densities derived from idealized coronal geometry. Departures from these assumptions, such as non-Kolmogorov spectra or asymmetric density distributions, may introduce additional uncertainties in velocity estimation. Furthermore, radio occultation techniques are inherently limited by LOS integration, which averages fluctuations across extended radial paths, potentially masking fine-scale spatial variations. Future studies involving longer duration campaigns, multi-frequency tracking, and improved turbulence diagnostics could help address these challenges and provide more precise constraints on coronal heating and wind acceleration mechanisms.

\section{Conclusions}
\label{sec:ANSWER}

In this study, we demonstrate the versatility of our technique by applying it to two contrasting observational periods: one characterized by low solar activity and slow solar wind speeds, and another associated with a coronal hole region exhibiting enhanced activity and fast solar wind. Although the spectral width method was originally developed for S-band radio signals, it remains applicable to X-band signals, such as those from the Akatsuki probe, after appropriate adjustments for the 2016 slow solar wind conditions \citep{Aggarwal2025}. For the 2022 observations near the coronal hole regions, a scaling factor was applied to the electron density estimates, highlighting the strong dependence of solar wind speed estimates on the local plasma density. Our analysis employed Eq. (\ref{eq:slowwind}), introduced as a general framework for deriving slow solar wind speeds from spectral broadening and geometrical considerations. The solar wind velocities derived from this radio occultation study align well with previously reported values (see Figs. \ref{fig:chiba} and \ref{fig:history}), clearly illustrating both the acceleration of the solar wind in the coronal region and the significant role of coronal holes in producing high-speed streams.

\section*{Acknowledgments}
The valuable inputs provided by the reviewers, Dr. Jason Kooi and Mx. Kenny Kenny, and the suggested references were deeply valued, leading to significant enhancements in the article's quality.
We would like to thank the Akatsuki mission team for monitoring the Akatsuki radio signals. The assistance provided by Dr. Keshav R. Tripathi from the University of Tokyo is greatly appreciated by the authors. The Prime Minister's Research scholarship (PMRF) program, Ministry of Education, Government of India, awarded author KA a research scholarship (PMRF-2103356).
The authors KA and AD acknowledge the use of facilities procured through the funding via the Department of Science and Technology, Government of India sponsored DST-FIST grant no. SR/FST/PSII/2021/162 (C) awarded to the DAASE, IIT Indore.

\section*{Data Availability}
Solar occultation and SPICE Kernel data from Akatsuki probe obtained from the Akatsuki Radio Science (RS) Data Archive at \href{https://darts.isas.jaxa.jp/planet/project/akatsuki/index.html.en}{https://darts.isas.jaxa.jp/planet/project/akatsuki/index} was used for the experiment. Position maps were created using the NASA SPICE toolkit. SDO/AIA images courtesy of NASA/SDO and the AIA, EVE, and HMI science teams (\href{https://sdo.gsfc.nasa.gov/data/}{https://sdo.gsfc.nasa.gov/data/}).

\bibliographystyle{mnras}
\bibliography{paper2} 
\bsp	
\label{lastpage}
\end{document}